# Playing with Black Strings[1]

Gary T. Horowitz

*Department of Physics, UCSB, Santa Barbara, CA. 93106*

**Abstract**

We review recent work showing that there exists a large class of new stable black strings which are not translationally invariant. Both neutral and charged black strings are considered. The discussion includes known properties of these new solutions, attempts to find them explicitly, and a list of open questions.

---

# 1 Introduction

It is a real pleasure to help celebrate Stephen Hawking's $60^{th}$ birthday. I have known Stephen for more than 20 years and had the pleasure of collaborating with him on three projects:

1) Positive mass theorems for black holes [1]
2) Entropy, area, and black hole pairs [2]
3) The gravitational Hamiltonian, action, entropy and surface terms [3]

The first arose when I was a postdoc at the Institute for Advanced Study at Princeton, and Stephen came to visit. We soon discovered that Malcolm Perry and I were working on the same problem that Stephen and Gary Gibbons were: How to generalize Witten's proof of the positive mass theorem [4] to include inner boundaries associated with black holes. We decided to join forces. I think it was because of this collaboration that Stephen invited me to visit Cambridge. I accepted and spent six weeks there in 1983. The thing I remember most about that visit was one day at tea, Stephen started to recite the names of the past Lucasian Professors of Mathematics starting with Newton. After a few names I expected him to stop. But he kept on going. He knew them all, up to the present.

I was thinking about what I could give Stephen for his birthday. Given his longstanding interest in black holes, I decided that the best present would be a new result about black holes. Since Stephen (and his colleagues) pretty much cleaned up the subject in four spacetime dimensions back in the 1970's [5], I decided to look in higher dimensions. (Motivated by string theory, there has been considerable interest in the properties of black holes in higher dimensions.) I was greatly assisted in this work by my postdoc Kengo Maeda. We found something quite surprising. In fact, I was stunned when we first obtained these results. Something that I thought was true for almost ten years turns out not to be.

As you know, ordinary black holes are stable. If you perturb the Schwarzschild solution, the perturbation remains small. In fact, it either radiates to infinity or falls into the black hole and decays at a certain characteristic rate. In higher dimensions, things are not so simple. We will consider the case of one extra dimension (which is compactified to a circle of length $L$ at infinity), but similar results hold in higher dimensions as well. In five spacetime dimensions, there is a solution which is just the product of a four dimensional

Schwarzschild black hole and a circle:

$$ds^2 = -\left(1 - \frac{r_0}{r}\right)dt^2 + \left(1 - \frac{r_0}{r}\right)^{-1} dr^2 + r^2 d\Omega + dz^2 \tag{1}$$

This looks like a one dimensional extended object surrounded by a horizon, and is characterized by two parameters, the Schwarzschild radius $r_0$ and $L$. In general, we will call any object with an event horizon having topology $S^2 \times S^1$ a "black string". A "black hole" will refer to an object with horizon topology $S^3$. (The most general topology consistent with a static horizon in five dimensions is a finite connected sum of $S^2 \times S^1$'s and (homotopy) $S^3$'s [6].) So (1) describes a black string. Following a suggestion by Hawking's student, Brian Whitt, Gregory and Laflamme [7] showed that this spacetime is unstable to linearized perturbations with a long wavelength along the circle. More precisely, there is a critical size $L_0$ of order $r_0$ such that black strings with $L \leq L_0$ are stable and those with $L > L_0$ are unstable. The unstable mode is spherically symmetric, but causes the horizon to oscillate in the $z$ direction. Gregory and Laflamme also compared the total entropy of the black string with that of a five dimensional black hole with the same total mass, and found that when $L > L_0$, the black hole had greater entropy. They thus suggested that the full nonlinear evolution of the instability would result in the black string breaking up into separate black holes which would then coalesce into a single black hole. Classically, horizons cannot bifurcate, but the idea was that under classical evolution, the event horizon would pinch off and become singular. When the curvature became large enough, it was plausible that quantum effects would smooth out the transition between the black string and black holes.

This idea that long black strings will break up into black holes has been widely accepted and assumed in various arguments. However I will try to convince you that this widespread belief is incorrect: Black strings do not in fact, break up into black holes. Under very weak assumptions, one can prove that an event horizon cannot pinch off in finite time. In particular, if one perturbs (1), an $S^2$ on the horizon cannot shrink to zero size in finite affine parameter. The basic idea is the following. Hawking's famous area theorem [8] is based on a local result that the divergence $\theta$ of the null geodesic generators of the horizon cannot become negative, i.e., the null geodesics cannot start to converge. If an $S^2$ on the horizon tries to shrink to zero size, the null geodesics on that $S^2$ must be converging. The total $\theta$ can stay positive only if the horizon is expanding rapidly in the circle direction. But this produces a large shear. If the $S^2$ were to shrink to zero size in finite time, one can show this shear would drive $\theta$ negative. A more physical argument shows that the horizon cannot slowly pinch off, taking an infinite time to do so. The net result is that the solution must settle down to a new (as yet unknown) static black string solution which is not translationally invariant along the circle.

One can view this result as an example of spontaneous symmetry breaking in general relativity. The most symmetric solution is unstable, and the stable solution has less symmetry. Unlike the usual particle physics examples where the broken symmetry is an internal one, here the broken symmetry is spatial translations.

Once one knows that stable black strings exist, one can use them as a tool to construct a larger class of non-uniform black strings which are unrelated to any instability. For example, if one considers Einstein Maxwell theory in five dimensions, there are translationally invariant, near extremal charged black strings which are believed to be stable. Nevertheless, we will see that there are also inhomogeneous charged black strings with the same mass and charge. In fact, the non-uniform solutions have much greater entropy. Unlike the neutral solutions, these new near extremal solutions can exist even when the size of the circle at infinity is very small.

In the next section we present the argument that stable non-uniform black strings must exist. We also discuss attempts to find these new solutions. In section three, we describe some properties of the new solutions, including the transition from uniform to non-uniform configurations. In the following section we generalize to charged black strings and show that there are non-uniform solutions which are unrelated to any instability. The fifth section contains a list of open questions and future directions. We end with a brief conclusion.

## 2 Existence of New (Vacuum) Solutions

We first prove that horizons cannot pinch off in finite proper time [9]. Consider initial data consisting of a constant $t$ surface in (1) with a spherically symmetric perturbation. We set $L > L_0$, and give the perturbation a long enough wavelength around the circle so that it grows exponentially in the linearized approximation. Now consider the full nonlinear evolution of this initial data using Einstein's vacuum field equation. We will assume that naked singularities do not form away from the horizon. This would be a blatant violation of cosmic censorship in five dimensions. It would be a much more interesting conclusion than the one we will find, but also much less likely. However, we will certainly allow the possibility that singularities form on the horizon. Our goal will be to show that this cannot happen.

Since the initial data is asymptotically flat, the maximal evolution should contain at least part of future null infinity $\mathcal{I}^+$. Since the initial data contains trapped surfaces which cannot lie in the past of $\mathcal{I}^+$, there must be an event horizon. This event horizon is a four dimensional null surface ruled by a family of null geodesics. Let $\lambda$ be an affine parameter along these geodesics and set $\ell^\mu = (\partial/\partial\lambda)^\mu$. Since we have assumed that the initial data is spherically symmetric, the same will be true for the evolved spacetime. Thus, the metric on a cross-section of the horizon at constant $\lambda$ can be written

$$ds^2 = e^{2\chi}dz^2 + e^{2\psi}d\Omega \qquad (2)$$

where $z$ is a coordinate along the $S^1$, and $\chi, \psi$ are functions of $\lambda$ and $z$. Notice that if the sphere shrinks to zero size, $\psi$ must go to minus infinity. The divergence of the null geodesic generators is

$$\theta = \dot\chi + 2\dot\psi \qquad (3)$$

where a dot denotes derivative with respect to $\lambda$. The change in the divergence along the null geodesics is given by the Raychaudhuri equation [5]. In five dimensions, this is

$$\dot\theta = -\frac{\theta^2}{3} - \sigma^{\mu\nu}\sigma_{\mu\nu} - R_{\mu\nu}\ell^\mu\ell^\nu \qquad (4)$$

where $\sigma_{\mu\nu}$ denotes the shear of the null geodesic congruence. This is a measure of how distorted a small sphere becomes when evolved along the null geodesics. For the metric (2) one finds

$$\sigma_{\mu\nu}\sigma^{\mu\nu} = \frac{2}{3}(\dot\chi - \dot\psi)^2 \qquad (5)$$

Since $\theta \geq 0$ and $\dot\psi \leq 0$, we have $\dot\chi \geq -2\dot\psi$ from (3) and hence $\sigma_{\mu\nu}\sigma^{\mu\nu} \geq 6\dot\psi^2$. Since the spacetime is Ricci flat, from (4) we have $\dot\theta \leq -6\dot\psi^2$. Thus if $\theta_0 > 0$ is the initial value of the divergence,

$$\theta(\lambda) \leq \theta_0 - 6\int_0^\lambda \dot\psi^2 \qquad (6)$$

Using $(\dot\psi + 1)^2 \geq 0$, this implies $\theta(\lambda) \leq 12\psi(\lambda) + 6\lambda+$ constant. It is now clear that if the sphere pinches off in finite affine parameter ($\psi \to -\infty$), then $\theta$ must become negative, and in fact go to minus infinity.

This leads to a contradiction as follows [10]. If $\theta < 0$ at some point $p$ on the event horizon, one can deform a cross-section of the horizon through $p$, so that it enters the past of future null infinity and still has the outgoing null geodesics converging. This is like having a trapped surface in the past of null infinity, and can be ruled out in the same way. One considers the boundary of the future of the deformed cross-section $T$. This must intersect future null infinity, but (being the boundary of a future set) through every point there must exist a past directed null geodesic which stays on the boundary until it reaches $T$. This is impossible since $\theta < 0$ on $T$ implies that all outgoing null geodesics have conjugate points[2], and cannot stay on the boundary of the future of $T$. (Notice that even though Hawking's original argument that $\theta > 0$ [8] assumed that the null geodesic generators of the horizon were complete, this is not required in the later proofs [10].) This completes the proof that the horizon cannot pinch off in finite time.

This result can be extended in many directions. Since it is based on a local calculation, one can apply it to nonspherical perturbations, higher dimensional spacetimes, horizons extended in more than one direction (black branes), and collapsing surfaces of various dimensions. In fact, one can show that no circle on the horizon can shrink down to zero size in finite affine parameter [9].

We now argue that it is unlikely that the horizon will pinch off even in infinite affine parameter. How could this happen? Fix some late time, and consider the geometry of the horizon near its smallest cross-section. The horizon cannot stay small for a distance much larger than its width, since in that case, it would resemble a thin black string and would be unstable. One does not expect a generic perturbation to approach an unstable solution at late time. But if the horizon remains small for only a distance of order its width, it will look like two spherical black holes connected by a small "neck". But the spacetime near the neck would be analogous to that obtained by bringing two black holes close together, and in that case it is well known that the horizon does not form a small neck. Instead, a new trapped surface forms which surrounds both black holes. This is simply because there is now double the mass within a sphere containing both black holes so the effective Schwarzschild radius moves out. Similarly, we would expect that if the apparent horizon tried to pinch off in infinite affine parameter by forming a small neck, there would be another apparent horizon form outside, and the true event horizon would not pinch off.

---

[2]A conjugate point is a point where the geodesic is crossed by a nearby null geodesic originating from $T$.

One might wonder it the final solution could end up independent of $z$, but with the proper size of the circle shrinking as one comes in from infinity, so that the horizon is stable. This is easily ruled out as follows: If the solution were independent of $z$, we could use a Kaluza-Klein reduction and view it as a static four dimensional black hole coupled to $\chi$, which acts like a massless scalar field. The usual "no hair" theorems show that $\chi$ must be constant. So the size of the fifth dimension must be constant if the black string is translationally invariant.

Given that the horizon cannot pinch off, one can ask what the solution will approach at late time. Any oscillatory motion will probably lose energy through gravitational radiation and damp out. The most plausible outcome is that the solution settles down to a new static black string which is not translationally invariant.[3] This is similar to the standard assumption in four dimensions, that solutions with a black hole will settle down at late time to Schwarzschild (or Kerr if there is nonzero angular momentum).

What do we expect the final state to look like? If $L$ is slightly larger than $L_0$, the solution will probably have a horizon with one maximum $S^2$ cross-section and one minimum. If $L \gg L_0$, there are two possibilities, depending on whether the size of the circle at the horizon shrinks to a value much smaller than $L$. (Since the solution is no longer translationally invariant, the above Kaluza-Klein reduction argument does not apply. Even though the circle can shrink, its size cannot go to zero without violating $\theta > 0$.) If the size of the circle shrinks, then one can again have a horizon with one maximum and one minimum cross-section. If the size of the circle does not shrink appreciably, it is likely that the local shape of the horizon is determined by $L_0$. This is because $L_0$ sets the scale between stable and unstable modes. We expect more or less regular oscillations on the horizon with the radii of maximum and minimum $S^2$'s, $r_{max}$ and $r_{min}$, differing from the initial Schwarzschild radius $r_0$ (which is of order $L_0$) by factors of order unity. $L$ just imposes periodicity at a long scale, but should not affect the local structure.

---

[3]The Weyl tensor on the horizon could, in principle, diverge even though $\theta > 0$ and the horizon has not pinched off [11]. This is because $\theta$ involves only first derivatives of the metric while the Weyl tensor involves second derivatives. But there is no reason to expect that such weak curvature singularities will form from generic perturbations of the original uniform black string.

Clearly, one would like to find these new solutions and there are several methods one might try. Since they are expected to be static and spherically symmetric in five dimensions, one essentially has a set of coupled nonlinear PDE's in two variables, $(r, z)$. These equations appear to be difficult to solve explicitly, but it is possible that a clever choice of coordinates will simplify the problem. (For one attempt in this direction, see [12]). After all, in four dimensions, the analogous problem of finding all static axisymmetric vacuum solutions can be solved completely. These are the Weyl metrics[4]. If an analytic solution cannot be found, one can always try to solve the equations numerically. This is similar to a two dimensional elliptic problem.

Another approach is to numerically solve the time dependent equations starting with initial data corresponding to a slightly perturbed uniform black string. This is currently underway [14]. Although they have not been able to reliably evolve long enough to see the final state, their results are already quite interesting. A dimensionless measure of the inhomogeneity of the black string is the ratio of the largest radius $S^2$ cross-section of the horizon to the smallest, $r_{max}/r_{min}$. In the numerical evolution, this ratio grows rapidly to about ten. It then starts slowing down and appears to stop growing at about thirteen. There is evidence that this is not yet the final static configuration, but at the moment, further evolution is unreliable due to numerical error. Work is continuing, but it is already clear that the horizon does not simply pinch off. One advantage of this evolution approach is that once the final state is found, it is guaranteed to be stable. One never numerically evolves to an unstable configuration without fine tuning.

## 3 Properties of the New Solutions

Even without explicit knowledge of the new solutions, one can deduce some of their properties. For example, the new solutions must approach the uniform black string exponentially fast at infinity. This is because at large distances from the horizon, the inhomogeneity should be washed out. It will thus resemble a perturbation of (1). Since $z$ is periodic, any $z$-dependent perturbation satisfies a massive spin two equation and must fall off exponentially.

---

[4]Unfortunately, a straightforward generalization of the Weyl ansatz does not lead to the same simplifications in five dimensions due to the fact that two-spheres have curvature while circles do not [13].

The surface gravity $\kappa$ must be constant over the static horizon, even when it is inhomogeneous. In [9] it was argued that for the static solution, the surface gravity, mass $\hat{M}$, and horizon area $A$ must satisfy

$$\hat{M} = \frac{\kappa A}{4\pi} \qquad (7)$$

The reason was the following. If $\xi$ denotes the static Killing field, then $*d\xi$ is closed by the vacuum Einstein equation. Consider the integral of $d*d\xi$ over a static slice from the horizon to infinity. The surface term at infinity yields the total mass $\hat{M}$, and the surface term at the horizon yields $\kappa A/4\pi$. Since these must be equal we obtain (7). However this formula cannot hold for all static solutions with horizons[5]. The first law of black hole mechanics [15] (which holds in all dimensions) implies $dM = \kappa dA/8\pi$, so (if $M = \hat{M}$) we can solve for $\kappa$ and obtain

$$M = 2A\frac{dM}{dA} \qquad (8)$$

or $A \propto M^2$. This is true for Schwarzschild in four dimensions but not in higher dimensions. The problem is that the surface integral at infinity of $*d\xi$ is the Komar mass and the mass that appears in the first law is the ADM mass. In four dimensions these two masses are the same but this is not true in higher dimensions. A simple counterexample is the product of time and the Euclidean Schwarzschild solution. This Ricci flat metric has nonzero ADM mass, but since $\xi^\mu = (\partial/\partial t)^\mu$ is covariantly constant, the Komar mass is zero. More generally, if $g_{zz}$ has a nonzero $1/r$ contribution, the Komar mass and ADM mass will differ. In short, (7) will hold for all static black strings provided $\hat{M}$ represents the Komar mass.

To understand the relation between the new solutions and the original translationally invariant ones, it is convenient to introduce two dimensionless variables. In five dimensions, the ADM mass has dimensions of length squared, so $M/L^2$ is dimensionless, where (as usual) $L$ is the length of the circle at infinity. This describes the overall shape of the black string. If it is large, the black string is short and fat, if it is small, the horizon is long and thin. The second parameter is the dimensionless measure of the inhomogeneity, $r_{max}/r_{min}$. This parameter is clearly never less than one, and when it is equal to one, there is a solution for every value of $M/L^2$ given by (1). (We previously described these solutions in terms of the two dimensionful parameters $r_0, L$, but to avoid trivial constant rescalings of the metric, it is convenient to use this dimensionless parameterization.)

---

[5]I thank S. Gubser for pointing out this problem.

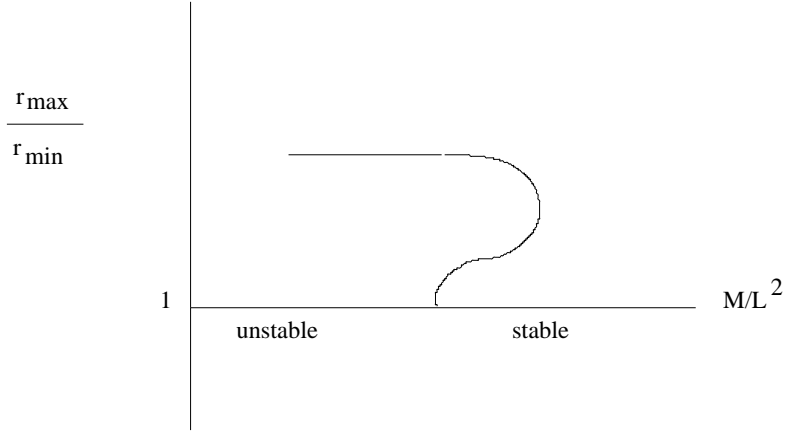

Figure 1: Near the transition where the uniform black string becomes unstable, the new static solutions probably behave as shown by the curve. This is supported by both a perturbative calculation and numerical evolution.

There is a critical value of $M/L^2$ of order one which separates the stable and unstable uniform black strings. At this critical value, there is a nontrivial static perturbation to (1) [7]. This strongly suggests that the new inhomogeneous black strings meet the old solutions at this point. One's first thought is that the new solutions will branch off to smaller values of $M/L^2$ since that is where the instabilities guarantee that new solutions must exist. Surprisingly, that is not what happens. The actual behavior is shown in the figure. The new solutions actually overlap part of the region of parameters where the uniform black string is stable. This was first found in a perturbative analysis near the critical point by Gubser [16]. He solved Einstein's equation to second order starting with the known static perturbation, and found that the inhomogeneous solution starts to increase $M/L^2$. This was later supported by numerical work. Starting with initial data with $M/L^2$ slightly below the critical value, the numerical evolution does not settle down to a slightly inhomogeneous black string [14]. Instead the inhomogeneities grow significantly, just as for smaller values of $M/L^2$. The main difference between the evolutions is just that when $M/L^2$ is slightly below the the critical value, the perturbation grows much more slowly. But eventually, the solution becomes just as inhomogeneous.

This indicates that the transition between the stable uniform black strings at large $M/L^2$ and stable inhomogeneous solutions at small $M/L^2$ involves a discontinuous jump in the horizon geometry. It is thus analogous to a first order phase transition. It also shows that there can be many solutions with the same value of $M/L^2$. Of the three solutions with $M/L^2$ slightly above the critical point, the middle one is expected to be unstable, and the other two are presumably stable. It is clear from another standpoint, that many solutions should exist. Given a stable and static black string with mass $M$ and length $L$, one can always "unwrap" it to obtain solutions with mass $nM$ and length $nL$. So every solution gives rise to series of static solutions with $M/L^2$ reduced by $1/n$ and $r_{max}/r_{min}$ unchanged. But the solutions with $n > 1$ need not be stable. There may be unstable modes with wavelength $L < \lambda < nL$. (This is just what happens if one starts with the stable uniform solution and unwraps it.) The existence of these different solutions with the same $M$ and $L$ show that the four dimensional no-hair theorems do not extend to higher dimensions, at least if one direction is compactified.

It has recently been shown that the no-hair theorems do not apply even if space is uncompactified [17]. One can take a black string and connect its ends to form a circle. This configuration would normally collapse down and form a spherical black hole, but it has recently been shown that one can stabilize it by adding rotation. In fact, Emparan and Reall [17] have found an explicit stationary axisymmetric vacuum solution in five dimensions describing this situation. It is parameterized by its total mass $M$ and angular momentum $J$. It is not yet known if these solutions are stable. For large $J$, the black string is long and thin and is likely to be subject to the Gregory-Laflamme instability. But a rotating inhomogeneous horizon will probably radiate energy and cause the radius of the circle to shrink. When the black string is short and fat, it might be stable. For some choices of $M$ and $J$ there is also a rotating black hole with the same parameters. This clearly shows that the no-hair theorem does not hold in higher dimensions.

# 4 New Charged Black Strings

So far we have talked about neutral black strings. If one adds a little charge not much changes. But near the extremal limit, one finds new phenomena. Gubser and Mitra [18] have conjectured that for a black string with a (noncompact) translational symmetry, there exists a Gregory-Laflamme instability if and only if the specific heat is negative. In other words, there is a close connection between classical stability and thermodynamic stability. Reall [19] has provided strong support for this conjecture (which was further studied in [20, 21]). Since the specific heat is often positive near extremality, one might expect:

1) There are no inhomogeneous black strings near extremality.
Also, since neutral black strings are unstable only if $M/L^2$ is small, one might expect

2) There are no inhomogeneous black strings with $M/L^2 \gg 1$.

We will see that both of the expectations are incorrect [22]. There exist stable inhomogeneous near extremal black strings, even when the uniform solution is stable. In other words, there are solutions with inhomogeneous horizons which are not just the result of an instability in the uniform solution. Furthermore, some of these solutions have arbitrarily large $M/L^2$. As an added bonus, we will see that there exist finite energy initial data with an apparent horizon of infinite area.

The simplest context to describe the new solutions is five dimensional Einstein-Maxwell theory with action

$$S = \int \sqrt{-g} \left[ \frac{R}{16\pi G} - \frac{1}{4} F_{\mu\nu} F^{\mu\nu} \right] d^5x, \tag{9}$$

where $G$ is the five dimensional Newton constant, $R$ is the scalar curvature, and $F$ is the Maxwell field. This theory is known to have both electrically charged black holes and translationally invariant, electrically charged black strings. The black holes are described by the five dimensional generalization of the Reissner-Nordstrom solution. We are mostly interested in the extremal limit which can easily be obtained as follows [23]. Let

$$ds^2 = -U^{-2}(x^i)dt^2 + U(x^i)\delta_{jk}dx^j dx^k \tag{10}$$

$$E_i = \alpha U^{-1} \partial_i U$$

where $\alpha = \pm(3/16\pi G)^{1/2}$. Then the Einstein-Maxwell equations reduce to just the condition that $U$ be a harmonic function. If $U$ is the field of a point mass,

$$U = 1 + \frac{\mu}{x_i x^i} \tag{11}$$

where $\mu$ is a positive constant, then the solution (10) describes an extremal black hole. The spatial metric has an infinite "throat" since the proper distance to $x^i = 0$ is stretched out infinitely, and near the origin the area of the three-spheres of constant radius is almost independent of the radius. In these coordinates, the event horizon is at $x^i = 0$ and has area

$$A_{\rm BH} = 2\pi^2 \mu^{3/2}. \tag{12}$$

The ADM mass $M$ and charge $Q$ are given by

$$M = \frac{3\pi}{4G}\mu, \qquad Q = \pm\sqrt{\frac{3\pi}{G}}\pi\mu, \tag{13}$$

respectively, where we have simply normalized the charge by $Q = \oint E_i dS^i$.

Solutions describing several extremal black holes are easily constructed by letting $U$ have several point sources. To compactify one direction, we let $U$ be the field of a one dimensional periodic array of point masses. The resulting metric can be written [23]

$$\begin{aligned} ds_5^2 &= -U^{-2}(r,z)dt^2 + U(r,z)(dr^2 + r^2 d\Omega^2 + dz^2), \\ U(r,z) &= 1 + \frac{\pi\mu}{Lr}\frac{\sinh 2\pi\frac{r}{L}}{\cosh 2\pi\frac{r}{L} - \cos 2\pi\frac{z}{L}}, \end{aligned} \tag{14}$$

where the coordinate $z$ is periodic with period $L$. The black hole horizon is located at $r = z = 0$, where $U$ diverges. Expanding $U$ near this point yields

$$U(r,z) = 1 + \frac{\mu}{r^2 + z^2} + \frac{\pi^2}{3}\frac{\mu}{L^2} + O\left(\frac{r^2}{L^2},\frac{z^2}{L^2}\right). \tag{15}$$

So the geometry near the horizon reduces to that of the isolated black hole. For $r \gg L$, we have

$$U = 1 + \frac{\pi\mu}{rL} + O(e^{-2\pi r/L}) \tag{16}$$

Note that the inhomogeneity in the $z$ direction falls off exponentially for large $r$ as expected. The ADM mass $M$ and charge $Q$ of the compactified solution are identical with that of the single black hole (13).

It is interesting to note that $\mu$ and $L$ are independent parameters in this solution: One can fit an arbitrarily large charged black hole into a space with one direction compactified on an arbitrarily small circle (at infinity). This is possible since the size of the circle depends on $U$. It follows from (16) that when $r \sim L$ the proper length of the circle is of order $\mu^{1/2}$, independent of $L$.

This theory also has the charged black string solution [22]:

$$ds_5^2 = -\left(1 - \frac{r_+}{r}\right)\left(1 - \frac{r_-}{r}\right) dt^2 + \left(1 - \frac{r_+}{r}\right)^{-1} dr^2 \tag{17}$$

$$+ r^2 \left(1 - \frac{r_-}{r}\right) d\Omega^2 + \left(1 - \frac{r_-}{r}\right)^{-1} dz^2.$$

$$F_{tr} = \pm \frac{1}{4r^2}\left(\frac{3r_+ r_-}{\pi G}\right)^{1/2} \tag{18}$$

The event horizon is at $r = r_+$ and has area

$$A = 4\pi r_+^2 L \left(1 - \frac{r_-}{r_+}\right)^{1/2}. \tag{19}$$

There is a curvature singularity at $r = r_-$. In the extremal limit, $r_+ \to r_-$, the horizon area clearly goes to zero. By a simple coordinate transformation, the extremal solution can be put into the form (10), where now $U$ is the field of a line source. The total mass $\tilde{M}$ and charge $\tilde{Q}$ is given by

$$\tilde{M} = \frac{L}{4G}(2r_+ + r_-), \qquad \tilde{Q} = \pm L\sqrt{\frac{3\pi r_+ r_-}{G}}. \tag{20}$$

One can check that near extremality, (17) has positive specific heat and is expected to be stable.

Equating the mass of the black hole to that of the black string in the extremal limit $r_+ = r_-$, we find $\mu = Lr_+/\pi$. It then follows from Eqs. (13) and (20) that the charges of the two systems are also equal. But we have seen that the horizon areas are very different.

The key question is: Can one add a small amount of mass to an extreme black hole in a space with one direction compactified, so that the horizon becomes topologically $S^2 \times S^1$? The answer is yes. The basic idea is to add a thin neutral black string that goes around the circle with its ends stuck on the black hole. The mass increases roughly by $r_0 L$ and for small $r_0$ this is much less than the total mass $M$, so the configuration remains near extremal. The area of the event horizon is still of order $M^{3/2}$. This configuration will evolve since the extreme black hole wants to swallow up the black string. But under evolution, the mass can only decrease since energy can be radiated away to infinity, and the horizon area can only increase. Since the horizon area is much greater than that of the uniform black string with the same mass and charge, it must settle down to a new inhomogeneous near extremal black string. This is true even though the uniform solution is stable.

To actually construct the initial data, it is easier to add the extreme black hole to the black string rather than vice versa. So we start with the spatial metric

$$ds_4^2 = u \left[ \left(1 - \frac{r_0}{r}\right)^{-1} dr^2 + r^2 d\Omega^2 + dz^2 \right] \qquad (21)$$

where $u$ is a function of $(r, z, \theta, \phi)$. $z$ is a periodic coordinate with period $L$. When $u = 1$, the metric represents the spatial part of the neutral black string solution with Schwarzschild radius $r = r_0$, while the $r_0 = 0$ case reproduces a spatial slice in the extremal solution (10). If we assume $E = \alpha u^{-1} \nabla u$ (where $\alpha$ is the same constant as before), and vanishing extrinsic curvature, the time symmetric constraint equations reduce to Laplace's equation on $u$. So we can let $u$ be the solution with point source at $r = r_0$, $\theta = 0$, and $z = 0$ representing a large charged black hole. Surfaces of constant $z \neq 0$ are wormholes like the maximally extended Schwarzschild solution, and the initial data (21) is invariant under reflections about $r = r_0$. This shows that $r = r_0$ is an apparent horizon.

We now encounter a surprise: The area of this apparent horizon is infinite! This is because the horizon of the extreme black hole is infinitely far away in the static surface, and so the black string must be infinitely long to reach it. This does not contradict our statement that the total mass is close to that of the black hole since the mass of the black string is infinitely redshifted near the horizon. It also does not contradict our statement that the event horizon will have finite area since the event horizon will lie outside the apparent horizon and not go down the infinite throat. Nevertheless it is certainly surprising that a finite energy initial data set can contain infinite area apparent horizons.

What will the final configuration look like? A static horizon must have constant surface gravity. The initial data we have constructed has zero surface gravity on the black hole and large surface gravity on the black string. But under evolution, we expect the black hole to swallow part of the black string and become slightly nonextremal. It is also plausible that the thin black string will acquire some charge and become near extremal. So the final configuration will probably resemble a large near extremal black hole with a near extremal black string going around the circle. This configuration would have low surface gravity everywhere. If one starts with this inhomogeneous black string and takes its extremal limit, it is plausible that the solution degenerates to an extremal black hole with a (singular) extremal black string going around the circle. This solution can be constructed explicitly. Recall the general form of static extremal solutions (10). The solution we want is obtained by letting $U$ be the harmonic function with a line source and a point source superposed on the line. The line source produces the extremal black string, while the point source reproduces the extremal black hole.

The idea of adding a thin black string to initial data containing a black hole can also be applied in the vacuum case without any charges. (This requires a generalization of the gluing construction [24] to rigorously obtain time symmetric initial data.) This is of particular interest when $M/L^2 \ll 1$. Consider the evolution of initial data describing a spherical black hole of mass almost $M$, with a very thin black string attached going around the circle. As before, the mass can only decrease, the horizon area can only increase, and the horizon topology must remain $S^2 \times S^1$. One might wonder what the final configuration now looks like, since the surface gravity must be constant. One possibility is that the proper size of the circle at the horizon shrinks significantly. The final black string will have entropy at least as large as the initial black hole. Let us compare this with initial data describing a perturbation of a long uniform black string with $M/L^2 \ll 1$. If the radius of the circle at the horizon does not shrink, this probably evolves into an inhomogeneous black string with horizon oscillations set by the initial Schwarzschild radius. This solution would have entropy much less than the black hole with the same mass.

# 5  Open Questions

**Geometry of event horizon**

(1) *What is the geometry of the horizon for the stable inhomogeneous black strings?*

This is clearly a key question. It cannot be answered just by looking at Einstein's equation near the horizon, since if there were a static configuration of matter outside, it would deform the horizon. One must impose the vacuum Einstein equation everywhere. A more specific question is:

(2) *Must the solution be spherically symmetric?*

As we discussed in section two, the proper size of the circle at the horizon might be quite different from $L$ when the solution is inhomogeneous. This leads us to ask:

(3) *Is the proper size of the circle at the horizon of order $L$, much larger, or much smaller?*

At the end of the previous section, we saw an example of a situation where it is likely that the size is much smaller.

**Transitions in the space of static solutions**

We have discussed the transition where the inhomogeneous black string meets the uniform solution. (See the discussion around Fig. 1) It is possible that there are further transitions at smaller values of $M/L^2$. This is because the new solutions with $M/L^2$ slightly below the critical value will likely have one maximum and one minimum radius of the horizon. As one decreases this parameter, there may be stable solutions with two local maxima and minima. This leads to the following question:

(4) *Are there static solutions with more than one maximum and minimum? If so, does the transition from one maximum to two proceed smoothly or discontinuously?*

For small values of $M/L^2$ there is another branch of solutions describing small five dimensional spherical black holes in a space with one direction compactified. As one increases $M/L^2$, it is intuitively clear that this branch becomes a black string.

(5) *What is the nature of the transition from black holes to black strings? Are the first black strings formed inhomogeneous? If so, do they join onto the previously discussed solutions?*

**Number of new solutions**

(6) *How many static black strings are there?*

We have been implicitly assuming that for each value of $M/L^2$ there are at most a finite number of solutions. But a simple Newtonian analogy suggests otherwise. In (five dimensional) Newtonian gravity, an $SO(3)$ invariant source is described by an arbitrary function $\rho(z)$ giving the line density on the axis. Of course, most of these configurations will not be static. But if one compactifies the $z$ direction on a circle, one expects both a uniform line density and a single point mass to be static. An arbitrary linear combination would provide a continuous one parameter family of solutions, all with the same $M/L^2$!

This argument certainly requires some modification in general relativity. Consider the case of $M/L^2 \gg 1$. If one imagines that the horizon is located where the Newtonian potential is of order one, one obtains a small inhomogeneity on a black string that is short and fat. But the corresponding translationally invariant solution is stable. One would expect a small perturbation to decay and not remain static. Nevertheless, it is possible that there is a range of small $M/L^2$ for which the number of static black strings is infinite.

In the case of extreme charged black strings, this Newtonian argument can be made more precise. From the form of the solutions (10) it is clear that there is a static solution for any choice of the line density $\rho(z)$. Of course these solutions do not have regular horizons. It is natural to ask:

(7) *Which of the solutions (10) with $U$ determined by an arbitrary line density $\rho(z)$ have nonextremal analogs with smooth event horizons?*

**Black string critical phenomena**

It is likely that there is a new type of black hole critical behavior in five dimensions. Recall the situation in four dimensions involving the gravitational collapse of a spherically symmetric massless scalar field. For any initial profile, if the initial amplitude is small, the scalar field will scatter as in flat spacetime. The curvature will remain small everywhere and there will be no singularities. If the initial amplitude is large, most of the scalar field collapses down to form a large black hole. Near a critical amplitude, which separates these two phases, one finds behavior similar to critical phenomena in condensed matter physics, including universal critical exponents [25].

This analysis of a spherically symmetric scalar field coupled to gravity in $D = 4$ can be immediately lifted to $D = 5$. This is because, if one assumes the vector potential vanishes, the standard Kaluza-Klein reduction of $D = 5$ vacuum Einstein theory is precisely Einstein gravity minimally coupled to a scalar field. More precisely, given any solution $\phi$, $g_{\mu\nu}$ of $S = \int \sqrt{-g} d^4x [R - 2(\nabla \phi)^2]$, then

$$ds^2 = e^{-4\phi/\sqrt{3}} dz^2 + e^{2\phi/\sqrt{3}} g_{\mu\nu} dx^\mu dx^\nu \qquad (22)$$

is a vacuum metric in five dimensions. Since this solution is independent of $z$, Choptuik's original analysis corresponds to studying a transition between nonsingular evolution and the formation of very thin black strings in $D = 5$. From the five dimensional viewpoint, this is very unnatural. One should clearly relax the assumption of translation invariance in the extra dimension. Near the transition point, any $z$ dependent perturbation of the initial data will probably result in the formation of small black holes (with $S^3$ topology) rather than a thin black string. The transition between nonsingular evolution and small spherical black holes in higher dimensions is similar to four dimensions [26] [6].

However, another type of transition should occur even at nonzero horizon area. One can easily construct initial data which evolves to a black hole with $S^3$ horizon. One can also construct initial data which evolves to a black string with $S^2 \times S^1$ horizon. But one can continuously interpolate between these initial configurations. Since each of the final states are expected to be stable, one might expect an open set of initial data evolve to each one. This leads to:

(8) *What happens at the transition point between the formation of black holes and black strings?*

As before, the transition can be reached by adjusting just one parameter. This dynamical question can be quite different from the static transition considered in question (5). After all, the critical solution in the four dimensional collapse [25] cannot be found by looking at the static Schwarzschild metrics.

**Generalizations**

One natural generalization (especially in light of recent observations) is to add a nonzero cosmological constant. It is known that de Sitter horizons can sometimes be unstable [27, 21]. The stability of black strings in anti-de Sitter spacetime has also been investigated [28].

(9) *What is the geometry of static inhomogeneous horizons when the cosmological constant is nonzero?*

We have been treating the black strings entirely classically. When quantum matter is included, black strings will Hawking radiate just like black holes. Even though the temperature is constant over the horizon, it is not clear what happens to the inhomogeneities:

(10) *Will $r_{max}/r_{min}$ increase, decrease, or remain the same under Hawking evaporation?*

---

[6]This was done without compactification, but near the transition point, the compactification should not have much affect.

# 6  Conclusions

Let me conclude by comparing static charged black holes in four and five dimensions (where the fifth dimension is compactified on a circle). Four dimensional black holes are rather boring. They always have topology $S^2$, they are stable, have maximum symmetry (spherically symmetric), and are characterized by just their charge and mass.

In five dimensions, things are more interesting. One has static charged black holes with topology $S^3$. These are direct analogs of the four dimensional case and are presumably stable and unique [29]. But in addition, there are black strings with topology $S^2 \times S^1$. These can be unstable, can be inhomogeneous, and can have more than one stable solution with the same mass and charge.

One has the impression that the results presented here may be just the tip of the iceberg. Higher dimensional generalizations of black holes seem to have a very rich structure that we are only beginning to explore.

Stephen has been an inspiration for me throughout my career. I wish him a very happy birthday!

### Acknowledgements

It is a pleasure to thank Kengo Maeda for collaboration on all aspects of the work described here. I am grateful to V. Hubeny for extensive discussion of the Gregory-Laflamme instability and the nature of the final state. I also thank S. Gubser, J. Isenberg, and S. Ross for useful discussions. Finally, I want to thank the organizers of the Future of Theoretical Physics and Cosmology conference, for a very stimulating meeting. This work was supported in part by NSF grant PHY-0070895.

# References

[1] G. W. Gibbons, S. W. Hawking, G. T. Horowitz and M. J. Perry, "Positive Mass Theorems For Black Holes," Commun. Math. Phys. **88** (1983) 295.

[2] S. W. Hawking, G. T. Horowitz and S. F. Ross, "Entropy, Area, and black hole pairs," Phys. Rev. D **51** (1995) 4302 [arXiv:gr-qc/9409013].

[3] S. W. Hawking and G. T. Horowitz, "The Gravitational Hamiltonian, action, entropy and surface terms," Class. Quant. Grav. **13** (1996) 1487 [arXiv:gr-qc/9501014].

[4] E. Witten, "A Simple Proof Of The Positive Energy Theorem," Commun. Math. Phys. **80** (1981) 381.

[5] S.W. Hawking and G.F.R. Ellis, *The large scale structure of spacetime*, Cambridge University Press (1973).

[6] M. l. Cai and G. J. Galloway, "On the topology and area of higher dimensional black holes," Class. Quant. Grav. **18** (2001) 2707 [arXiv:hep-th/0102149].

[7] R. Gregory and R. Laflamme, "Black Strings And P-Branes Are Unstable," Phys. Rev. Lett. **70** (1993) 2837 [arXiv:hep-th/9301052].

[8] S. W. Hawking, "Gravitational Radiation From Colliding Black Holes," Phys. Rev. Lett. **26** (1971) 1344.

[9] G. T. Horowitz and K. Maeda, "Fate of the black string instability," Phys. Rev. Lett. **87** (2001) 131301 [arXiv:hep-th/0105111].

[10] R. Wald, *General Relativity*, University of Chicago Press (1984).

[11] S. Hawking, private communication

[12] T. Harmark and N. Obers, "Black holes on cylinders," arXiv:hep-th/0204047.

[13] R. Emparan and H. S. Reall, "Generalized Weyl solutions," Phys. Rev. D **65** (2002) 084025 [arXiv:hep-th/0110258].

[14] M. Choptuik, L. Lehner, I. Olabarrieta, R. Petryk, F. Pretorius, and H. Villegas, to appear.

[15] R. M. Wald, "The First Law Of Black Hole Mechanics," arXiv:gr-qc/9305022.

[16] S. S. Gubser, "On non-uniform black branes," arXiv:hep-th/0110193.

[17] R. Emparan and H. S. Reall, "A rotating black ring in five dimensions," Phys. Rev. Lett. **88** (2002) 101101 [arXiv:hep-th/0110260].

[18] S. S. Gubser and I. Mitra, "Instability of charged black holes in anti-de Sitter space," arXiv:hep-th/0009126; "The evolution of unstable black holes in anti-de Sitter space," JHEP **0108** (2001) 018 [arXiv:hep-th/0011127].

[19] H. S. Reall, "Classical and thermodynamic stability of black branes," Phys. Rev. D **64** (2001) 044005 [arXiv:hep-th/0104071].

[20] J. P. Gregory and S. F. Ross, "Stability and the negative mode for Schwarzschild in a finite cavity," Phys. Rev. D **64** (2001) 124006 [arXiv:hep-th/0106220].

[21] V. E. Hubeny and M. Rangamani, "Unstable horizons," arXiv:hep-th/0202189.

[22] G. T. Horowitz and K. Maeda, "Inhomogeneous near-extremal black branes," arXiv:hep-th/0201241.

[23] R. C. Myers, "Higher Dimensional Black Holes In Compactified Space-Times," Phys. Rev. D **35**, 455 (1987).

[24] J. Isenberg, R. Mazzeo and D. Pollack, "Gluing and wormholes for the Einstein constraint equations," arXiv:gr-qc/0109045.

[25] M. W. Choptuik, "Universality And Scaling In Gravitational Collapse Of A Massless Scalar Field," Phys. Rev. Lett. **70** (1993) 9.

[26] D. Garfinkle, C. Cutler and G. C. Duncan, "Choptuik scaling in six dimensions," Phys. Rev. D **60** (1999) 104007 [arXiv:gr-qc/9908044].

[27] R. Bousso and S. W. Hawking, "(Anti-)evaporation of Schwarzschild-de Sitter black holes," Phys. Rev. D **57** (1998) 2436 [arXiv:hep-th/9709224].

[28] R. Gregory, "Black string instabilities in anti-de Sitter space," Class. Quant. Grav. **17** (2000) L125 [arXiv:hep-th/0004101]; T. Hirayama and G. Kang, "Stable black strings in anti-de Sitter space," Phys. Rev. D **64** (2001) 064010 [arXiv:hep-th/0104213].

[29] G. W. Gibbons, D. Ida and T. Shiromizu, "Uniqueness and non-uniqueness of static vacuum black holes in higher dimensions," arXiv:gr-qc/0203004.